\documentclass[a4paper]{jpconf}
\usepackage{graphicx}
\usepackage{amsmath}
\usepackage{slashed}
\begin{document}
\title{Anomalous Transport Properties of Dense QCD in a Magnetic Field}

\author{Vivian de la Incera}

\address{Dept. of Eng. Sci. and Physics, College of Staten Island,
CUNY, and CUNY-Graduate Center,
           New York 10314, USA}

\ead{vivian.incera@csi.cuny.edu}

\begin{abstract}
Despite recent advancements in the study and understanding of the phase diagram of strongly interacting matter, the region of high baryonic densities and low temperatures has remained difficult to reach in the lab. Things are expected to change with the planned HIC experiments at FAIR in Germany and NICA in Russia, which will open a window to the high-density-low-temperature segment of the QCD phase map, providing a unique opportunity to test the validity of model calculations that have predicted the formation of spatially inhomogeneous phases with broken chiral symmetry at intermediate-to-high densities. Such a density region is also especially relevant for the physics of neutron stars, as they have cores that can have several times the nuclear saturation density.  On the other hand, strong magnetic fields, whose presence is fairly common in HIC and in neutron stars, can affect the properties of these exotic phases and lead to signatures potentially observable in these two settings. In this paper, I examine the anomalous transport properties produced by the spectral asymmetry of the lowest Landau level (LLL) in a QCD-inspired NJL model with a background magnetic field that exhibits chiral symmetry breaking at high density via the formation of a Dual Chiral Density Wave (DCDW) condensate. It turns out that in this model the electromagnetic interactions are described by the axion electrodynamics equations and there is a dissipationless Hall current. 
\end{abstract}

\section{Introduction}
A growing body of works indicates that with increasing density, the chirally broken phase of quark matter is not necessarily replaced by an homogeneous, chirally restored phase, but instead, at least for an intermediate region of densities, the system may favor the formation of inhomogeneous phases. To gain insight on why this occurs notice that with increasing density the homogeneous chiral condensate becomes disfavored due to the high-energy cost of exciting the antiquarks from the Dirac sea to the Fermi surface where the pairs form. At the same time, with higher densities, co-moving quarks and holes at the Fermi surface may pair with minimal energy cost through a mechanism analogous to Overhauser's \cite{Overhauser}, giving rise to a spatially modulated chiral condensate \cite{InhCC}. Spatially modulated chiral condensates in QCD have been discussed in the context of quarkyonic matter \cite{QqM}, where they appear in the form of quarkyonic chiral spirals \cite{QCS} at zero magnetic field, or double quarkyonic chiral spirals  \cite{DQCS} in the presence of a magnetic field. Inhomogeneous chiral condensates have been also studied in the context of NJL models (for a review see \cite{InhCRev}) that share the chiral symmetries of QCD and are then useful to study the chiral phase transition. 

In the present paper, we explore the properties of a QCD-inspired NJL model with a background magnetic field that exhibits chiral symmetry breaking at high density via the formation of a Dual Chiral Density Wave (DCDW) condensate. As will be shown below, the electromagnetic interactions in this model are described by the equations of axion electrodynamics, exhibiting a dissipationless Hall current perpendicular to the magnetic field, and an anomalous electric charge directly proportional to the magnetic field strength and the modulation of the condensate.

\section{DCDW Phase in a Magnetic Field}

Let us describe cold and dense quark matter in a magnetic field with the help of a QCD-inspired the three-color, two-flavor Nambu-Jona-Lasinio (NJL) Lagrangian density that includes electromagnetism,
\begin{equation} \label{L_NJL_QED}
\mathcal{L}=-\frac{1}{4}F_{\mu\nu}F^{\mu\nu}+\bar{\psi}[i\gamma^{\mu}(\partial_\mu+iQA_{\mu})+\gamma_0 \mu]\psi 
+G[(\bar{\psi}\psi)^2+(\bar{\psi}i\mathbf{\tau}\gamma_5\psi)^2],
\end{equation}
Here $\psi^T=(u,d)$ is the fermion doublet in flavor space with electric charge matrix $Q=\mathrm{diag}(e_u,e_d)=\mathrm{diag}(\frac{2}{3}e,-\frac{1}{3}e)$; $\mu$ is the baryon chemical potential; G is the four-fermion coupling; and $\mathbf{\tau}_a$ are the isospin Pauli matrices. A constant and uniform magnetic field along the $x^3$-direction is introduced by the background field $\bar{A}_{\mu}=(0,0,Bx,0)$. Hence, the electromagnetic potential $A_{\mu}$ is the sum of the fluctuation field plus the background field $\bar{A}_{\mu}$. Notice that the flavor symmetry $SU(2)_L\times SU(2)_R$ of the NJL model is reduced to the subgroup $U(1)_L\times U(1)_R $ here due to the electromagnetic coupling. 
 
As shown in \cite{KlimenkoPRD82,PLB743}, this system favors the formation of the inhomogeneous condensate, 
\begin{equation}
\langle\bar{\psi} \psi\rangle= \Delta \cos(\mathbf{q} \cdot \mathbf{x}), \quad   \langle\bar{\psi}i \tau_3\gamma_5 \psi\rangle= \Delta \sin(\mathbf{q} \cdot \mathbf{x}),
\end{equation}
which represents a density wave with magnitude $\Delta$ and modulation vector $q_{i}=(0,0,q)$ along the field direction. This type of density wave condensate is known in the literature as the DCDW condensate \cite{PRD71}.  

The corresponding mean-field Lagrangian is
\begin{equation} \label{DCDW-MF_L}
\mathcal{L}_{MF}=\bar{\psi}[i\gamma^{\mu}(\partial_\mu+iQA_{\mu})+\gamma_0\mu]\psi
-m\bar{\psi}e^{i\tau_3\gamma_5qz}\psi-\frac{1}{4}F_{\mu\nu}F^{\mu\nu}- \frac{m^2}{4G},
\end{equation}
with $m=-2G\Delta$. The spatially dependent mass term can be eliminated with the help of the local chiral transformation 
\begin{eqnarray} \label{XTransf}
&\psi&\to U_A\psi \qquad \bar{\psi} \to \bar{\psi}\bar{U}_A
\\
\textrm{with} &U_A&=e^{-i\tau_3\gamma_5\theta} \quad \bar{U}_A=\gamma_0U^\dag\gamma_0=e^{-i\tau_3\gamma_5 \theta}
\end{eqnarray}

with $\theta(t,\mathbf{x})=\frac{1}{2}q_\mu x^\mu=\frac{qz}{2}$.

The mean-field Lagrangian then becomes
\begin{equation}\label{U_1-MF_L}
\mathcal{L}_{MF}=\bar{\psi}[i\gamma^{\mu}(\partial_\mu+iQA_{\mu}-i\tau_3\gamma_5\partial_{\mu}\theta) +\gamma_0\mu-m]\psi 
-\frac{1}{4}F_{\mu\nu}F^{\mu\nu}-\frac{m^2}{4G}
\end{equation}
The energy spectrum of the quarks in (\ref {U_1-MF_L}) separates into the LLL  ($l=0$) modes
\begin{equation}\label{LLLspectrum}
E^{LLL}_k=\epsilon\sqrt{\Delta^2+k_3^2}+q/2,  \quad \epsilon=\pm,
\end{equation}
and higher ($l>0$) Landau level modes
\begin{equation}\label{HighLspectrum}
E^{l>0}_k= \epsilon\sqrt{(\xi\sqrt{\Delta^2+k_3^2}+q/2)^2+2|e_fB|l}, \quad \epsilon=\pm, \xi=\pm, l=1,2,3,...
\end{equation} 

Notice that the LLL spectrum is not symmetric about the zero energy level \cite{KlimenkoPRD82}-\cite{PLB743}.  An asymmetric spectrum is a sign of a nontrivial topology. Indeed, it is known that such a theory is characterized by the topological quantity $\eta=\sum_k \mathrm{sgn}(E_{k})$, known as the Atiyah-Patodi-Singer (APS) invariant \cite{AS}. Henceforth, we use the notation $\eta_B$ for the APS invariant and keep in mind that due to the magnetic field $B$, the sum over the energy modes $E_k$ is equivalent to a sum over all the Landau levels and an integral over momenta. Notice that $\eta_B$ is divergent. In order to ensure that the energies with equal magnitude and opposite signs cancel out in the sum, it needs proper regularization. Using the regularization discussed in \cite{NS}, it can be shown that the contributions of all the $l>0$ modes cancel out because they are symmetric about zero, so only the asymmetric modes contribute to $\eta_B$, as was expected, given the connection between the nontrivial topology and the LLL spectral asymmetry. The regularized index $\eta_B=\lim_{s\to0}\sum_k \mathrm{sgn}(E_{k})|E_{k}|^{-s}$ gives rise to an anomalous baryon (quark) number density $ \rho^A_B$ \cite{PLB743} given by
 \begin{equation}\label{Baryon_charge}
 \rho^A_B=-N_c\eta_B/2=N_c\sum_{f} \frac{|e_f|}{4\pi^2}\mathbf{B} \cdot \mathbf{\bigtriangledown}(\mathbf{q}\cdot\mathbf{x})=3\frac{|e|}{4\pi^2}qB
 \end{equation}
when $q<2m$. The use of a different regularization procedure that allows to extract the anomalous part of the thermodynamic potential, led to the same anomalous quark number density, obtained in this case not from the index $\eta_B$, but as the derivative of the thermodynamic potential with respect to the baryon chemical potential $\mu$ \cite{KlimenkoPRD82}. The extension of this calculation to the isospin asymmetric case was done in \cite{PRD92}.
 When $q>2m$, the quark density acquires an additional, non-topological contribution \cite{PLB743} and becomes 
\begin{equation}
\rho^A_B=-N_c\eta_B/2=-N_c |eB|\left[-q+ \sqrt{q^2-4m^2}\right]/4\pi^2
 \end{equation}
 
As discussed in \cite{VI2016}, the lack of  invariance of the fermion measure in the path integral under the $U_A$ transformation produces an additional contribution to the Lagrangian density (\ref{U_1-MF_L}) given by
\begin{equation}
(\det U_A)^{-2}=e^{-2 \mathrm{Tr} \log U}=e^{i\int d^4x \theta(x)d(x)}
\end{equation}
where $d(x)=-2\delta^{(4)}(0) \mathrm{tr} \tau_3 \gamma_5$ and $\delta^{(4)}(0)= \langle x \mid x \rangle =\int \frac{d^4p}{(2\pi)^4}e^{i p(x-y)\mid_{x=y}}$. In the above equation, $\mathrm{Tr}$ means functional and matrix trace, while $\mathrm{tr}$ means matrix trace. This expression is ill-defined since it is the product of infinite ($\delta^{(4)}(0)$) times zero ($\mathrm{tr} \tau_3 \gamma_5=0$). It hence needs regularization and this should be done in a way that preserves the gauge-invariance of the theory. With that aim, we introduce an arbitrary, smooth regularizing function $f(t)$, such that $f(0)=1$, $f(\infty)=0$, $tf'(t)=0$ at $t=0$ and $t=\infty$. We then regularize the exponent following Fujikawa's approach \cite{FujikawaPRD21_1980} to obtain 
\begin{equation}\label{regularize}
\int_R d^4x \theta(x) d(x)=-2\lim_{\Lambda\to\infty}\mathrm{Tr} \left[ \theta(x) \tau_3 \gamma_5 f((i \slashed{D}/ \Lambda)^2)\right]=\int d^4x \frac{\kappa}{4}\theta F_{\mu\nu}\widetilde{F}^{\mu\nu}
 \end{equation}
with $D_\mu=\partial_\mu+iQA_{\mu}-i\tau_3\gamma_5\frac{q}{2}$ the corresponding Dirac operator, and $\kappa=\frac{N_c}{2\pi^2}[e_u^2-e_d^2]=\frac{e^2}{2\pi^2}$. 

The effective Lagrangian of the model is then
\begin{eqnarray} \label{Eff-MFL+axion}
\mathcal{L}_{eff}&=&\bar{\psi}[i\gamma^{\mu}(\partial_\mu+iQA_{\mu}-i\tau_3\gamma_5\frac{q}{2}) +\gamma_0\mu-m]\psi -\frac{m^2}{4G}
\nonumber
\\
&-&\frac{1}{4}F_{\mu\nu}F^{\mu\nu}+\frac{\kappa}{4}\theta F_{\mu\nu}\widetilde{F}^{\mu\nu},
\end{eqnarray}
Thus, the measure contribution generated an axion term $\frac{\kappa}{4}\theta F_{\mu\nu}\widetilde{F}^{\mu\nu}$ which, as will be seen below, leads to important new physics \cite{VI2016}. 

\section{Spectral Asymmetry and Index of the Dirac Operator}
The origin of the axion term in (\ref{Eff-MFL+axion}) can be traced back to the spectral asymmetry of the LLL. This can be better understood with the help of the index of the Dirac operator in Euclidean space
\begin{equation}
\textrm{index}(i \slashed{D}_E)=n^+-n^-,
\end{equation}
which is defined as the difference between the number of zero modes of $i\slashed{D}_E$ with positive $n^+$ and negative $n^-$ chirality. To see the connection, let us  consider

\begin{equation}\label{index}
\int d^4x d(x)=-2\lim_{\Lambda\to\infty}\mathrm{Tr} \left[ \tau_3 \gamma_5 f((i \slashed{D}_E/ \Lambda)^2)\right]=\int d^4x \frac{\kappa}{4}F_{\mu\nu}\widetilde{F}^{\mu\nu}
\end{equation}
which can be seen from (\ref{regularize}) after going to Euclidean space and taking $\theta(x)=1$. To keep the notation simple, we are not attaching an E index to the tensors and integral variables, but they are all in Euclidean.

Since $\tau_3$ and $i \slashed{D}_E$ are diagonal in flavor space, we can do our analysis separately for the u and the d quark and add the two results at the end. In Euclidean space the operator  $i \slashed{D}_E$ is hermitian, hence its eigenvalues $\lambda_k$ are real.  Consider a set of orthogonal and complete eigenfunctions $\phi_k$ of $i \slashed{D}_E$ with eigenvalues $\lambda_k$, $i \slashed{D}_E \phi_k=\lambda_k\phi_k$. Then, for each $\phi_k$, we have $i \slashed{D}_E \gamma_5\phi_k=-\lambda_k \gamma_5 \phi_k$, so $\gamma_5\phi_k$ is also an eigenfunction of $i \slashed{D}_E$ but with eigenvalue $-\lambda_k$. This implies that for $\lambda_k\neq 0$ the eigenfunctions $\phi_k$ and $\gamma_5 \phi_k$ are orthogonal and with opposite eigenvalues. It also follows that for $\lambda_k\neq 0$, we can construct two combinations of $\phi_k$ and $\gamma_5 \phi_k$,
\begin{equation}
\phi_{k,\pm}=\frac{1\pm\gamma_5}{2}\phi_k
\end{equation}
that are eigenfunctions of $\gamma_5$ and of $(i \slashed{D}_E)^2$, but not of $i \slashed{D}_E$.

We then have that if $\lambda_k\neq 0$ the functions $\phi_{k,\pm}$ satisfy 
\begin{equation}
 \gamma_5\phi_{k,\pm}=\pm\phi_{k,\pm}, \qquad (i \slashed{D}_E)^2\phi_{k,\pm}=\lambda_k^2 \phi_{k,\pm}
\end{equation}
meaning that for $\lambda_k\neq 0$ the eigenfunctions of $(i \slashed{D}_E)^2$ come in pairs with equal energy square eigenvalue, but opposite chirality.

On the other hand, for $\lambda_k=0$, both eigenfunctions $\phi_k$ and $\gamma_5\phi_k$ have the same zero eigenvalue under the operator $i \slashed{D}_E$. In the subspace of these zero-mode-eigenfunctions, which we denote by $\phi^{(0)}_l$,   $\gamma_5$ can be diagonalized, thus the $\phi^{(0)}_k$ can be separated into $\phi^{(0+)}_k$, with plus $\gamma_5$-eigenvalue (positive chirality), and $\phi^{(0-)}_k$, with minus $\gamma_5$-eigenvalue (negative chirality). However, there is no reason to claim that the numbers of eigenfunctions with positive and negative chiralities in the zero-mode-subspace have to be equal. In other words, the eigenfunctions $\phi^{(0\pm)}_k$ do not have to come in pairs.

Let us go back now to Eq. (\ref{index}), and consider the contribution coming from the u-quark. Taking the Tr operator in the space of the eigenfunctions $\phi^{\pm}$ of $(i\slashed{D}_E)^2$ we find

\begin{eqnarray} \label{indexdensity}
&-&\frac{1}{2}\int d^4x d^{(u)}(x) = \lim_{\Lambda\to\infty} \mathrm{Tr} \left[\gamma_5 f((i \slashed{D}_E/ \Lambda)^2)\right]
\nonumber
\\
&=&\lim_{\Lambda\to\infty}\sum_k\langle \phi_k \mid \gamma_5 f((i \slashed{D}_E/ \Lambda)^2) \mid \phi_k \rangle =\lim_{\Lambda\to\infty}\sum_k f(\lambda_k^2/ \Lambda^2) \langle \phi_k \mid \gamma_5 \mid \phi_k \rangle
\nonumber
\\
&=& \lim_{\Lambda\to\infty}\sum_{l=1}^{n^u+} f(0) \langle \phi^{(0+)}_l \mid \phi^{(0+)}_l\rangle
-\lim_{\Lambda\to\infty}\sum_{l=1}^{n^u_-} f(0) \langle \phi^{(0-)}_l  \mid \phi^{(0-)}_l\rangle 
\nonumber
\\
&=&\sum_{l=1}^{n^u+} \langle \phi^{(0+)}_l \mid \phi^{(0+)}_l\rangle - \sum_{l=1}^{n^u_-} \langle \phi^{(0-)}_l  \mid \phi^{(0-)}_l\rangle =n^u_+-n^u_-
\end{eqnarray} 
where we used that $f(0)=1$. In going from line 2 to 3 in Eq.(\ref{indexdensity}), we took into account that the eigenfunctions with nonzero $\lambda_k$ come in pairs of opposite chirality, so their contributions cancel out in the sum. This means that only the zero-energy modes contribute.

Doing the same calculation for the d quark and combining the two results we obtain
\begin{equation}
-\frac{1}{2}\int d^4x d(x)=n_+-n_-=\textrm{index}(i \slashed{D}_E)=-\int d^4x \frac{\kappa}{8}F_{\mu\nu}\widetilde{F}^{\mu\nu}
\end{equation}
so the integral of d(x) is proportional to the index of the Dirac operator and we can interpret d(x) as the index "density." The above relation remains valid also if the first and last terms are written back in Minkowski space.

 Now, we can see from the expressions (\ref{LLLspectrum})-(\ref{HighLspectrum}) in the massless limit case $\Delta=0$, that in the presence of a background magnetic field the higher Landau levels $(l>0)$ lead to no zero modes. Moreover, from the two asymmetric modes in the LLL, only one of them can be a zero mode. Therefore, the spectral asymmetry of the LLL gives rise to a nonzero $\textrm{index}(i \slashed{D}_E)$.  We conclude that the index "density" d(x) is connected to the spectral asymmetry of the LLL, and so is the axion term in (\ref{Eff-MFL+axion}). 

\section{Axial Electrodynamics in Dense QCD Matter in a Magnetic Field}
The electromagnetic effective action, found from the partition function after integrating in the fermion fields and expanding in powers of $A_{\mu}$, takes the form
\begin{eqnarray} \label{EA}
\Gamma(A)=&-V\Omega+\int d^4x \left[-\frac{1}{4}F_{\mu\nu}F^{\mu\nu}+\frac{\kappa}{4}\theta F_{\mu\nu}\widetilde{F}^{\mu\nu}\right]
\nonumber
\\
&-\int d^4x A^\mu(x) J_\mu(x)+\cdots,
\end{eqnarray}
with V the four-volume, $\Omega=\Omega(\mu,B)$ the thermodynamic potential \cite{KlimenkoPRD82}, and $J_\mu(x)=(J_0,\mathbf{J})$ the ordinary (nonaxion) electric four-current determined by the sum of the tadpole diagrams of each flavor. 

The linear equations of motion derived from (\ref{EA}) are
\begin{eqnarray}
&\mathbf{\nabla} \cdot \mathbf{E}=J_0+\frac{e^2}{4\pi^2}qB, \label{1}
 \\
&\nabla \times \mathbf{B}-\frac{\partial \mathbf{E}}{\partial t}=\mathbf{J}-\frac{e^2}{4\pi^2} \mathbf{q}\times \mathbf{E},  \label{2}
\\
&\mathbf{\nabla} \cdot \mathbf{B}=0, \quad \nabla \times \mathbf{E}
+\frac{\partial \mathbf{B}}{\partial t}=0 \label{3},
\end{eqnarray} 
These are the equations of axion electrodynamics. In the above equations, the terms $J_0$ and $\mathbf{J}$ represent the ordinary charge and current density contributions obtained from the tadpole diagrams, while  $J^{anom}_0=\frac{e^2}{4\pi^2}qB$ and  $\mathbf{J}^{anom}=-\frac{e^2}{4\pi^2} \mathbf{q}\times \mathbf{E}$ are anomalous charge and current densities generated by the axion term in (\ref{EA}). As shown in \cite{VI2016}, the tadpoles contributions do not cancel the anomalous ones.

As already discussed, the origin of the axion term in the effective action and thus the anomalous terms in Eqs. (\ref{1}-\ref{2}), can be traced back to the asymmetry of the LLL. This relation can also be seen from the connection between the anomalous electric charge and the anomalous quark number density associated with the Atiyah-Patodi-Singer index. Notice that the same expression for the anomalous electric charge density is found if one takes the anomalous quark number density of flavor f, $\rho^A_{Bf}=N_c\frac{|e_f|}{4\pi^2}qB$, multiplies it by the flavor's electric charge $e_f$, and sums in flavor. We highlight that the anomalous current is a dissipationless Hall current, perpendicular to both, the magnetic and the electric field. This could have important consequences for the transport properties of the system. Notice that no CME current is generated here because the axion field $\theta$ is time-independent. 

\section{Anomalous transport}

The magnetic DCDW phase exhibits quite interesting properties. The most important is the existence of the dissipationless anomalous Hall current, $\mathbf{J}^{anom}=-\frac{e^2}{4\pi^2} \mathbf{q}\times \mathbf{E}$.  Since it is perpendicular to $\mathbf{E}$ and to the condensate modulation $\mathbf{q}$, which in turn is parallel to $\mathbf{B}$, the Hall current is produced as long as $\mathbf{E}$ and $\mathbf{B}$ are not parallel.


These results can be relevant for neutron stars. Consider a neutron star with a core of DCDW matter threaded by a poloidal magnetic field. Any electric field present in the core, whether due to the anomalous electric charge or not, and as long as it is not parallel to the magnetic field, will lead to dissipationless Hall currents in the plane perpendicular to the magnetic field. Could these currents serve to resolve the issue of the stability \cite{Bstability} of the magnetic field strength in magnetars? How will the electric transport be affected by the anomalous Hall conductance? These and other questions highlight the importance to explore which observable signatures could be identified and then used as telltales of the presence of the DCDW phase in the core. Notice that the condition of electrical neutrality does not need to be satisfied locally for compact hybrid stars \cite{Glendenning}, which could have a core in the magnetic DCDW phase with an anomalous charge contribution and Hall currents circulating inside and at the surface.

The anomalous Hall current could be also produced in future HIC like those planned at the Nuclotron-based Ion Collider Facility (NICA) at Dubna, Russia \cite{PRC85} and at the Facility for Antiproton and Ion Research (FAIR) at Darmstadt, Germany \cite{1607.01487}, which will explore the high density, cold region of the QCD phase map, and where event-by-event off-central collisions will likely generate perpendicular electric and magnetic fields \cite{NICA}.  It will be interesting to carry out a detailed quantitative analysis of how these currents could lead to observable signatures, even after taking into account that there the QGP distributes itself more as an ellipsoid than as an sphere about the center of the collision. The Hall currents will tend to deviate the quarks from the natural outward direction from the collision center and one would expect a different geometry of the particle flow in the DCDW phase compared to other dense phases that have no anomalous electric current. The realization of the DCDW phase in the QGP of future HIC experiments is likely viable because the inhomogeneity of the phase is characterized by a length $\Delta x = \hbar / q \sim 0.6 fm$ for $q \sim \mu =300$ MeV \cite{KlimenkoPRD82}, much smaller than the characteristic scale $L\sim 10 fm$ of the QGP at RHIC, NICA,  and FAIR, while the time scale for this phase will be the same as for the QGP. 

Other interesting effects might emerge by considering the fluctuations $\delta\theta$ of the axion field. If one goes beyond the mean-field approximation, there will be mass and kinetic terms of the axion field fluctuation.  Besides, due to the background magnetic field, the axion fluctuation couples linearly to the electric field via the term $\kappa\delta\theta \mathbf{E}\cdot \mathbf{B}$, so the field equations of the axion fluctuation and the electromagnetic field will be mixed, giving rise to a quasiparticle mode known as the axion polariton mode \cite{axpolariton}. The axion polariton mode is gapped with a gap proportional to the background magnetic field. This implies that electromagnetic waves of certain frequencies will be attenuated by the DCDW matter, since in the DCDW medium they propagate as polaritons. The axion polariton could help to probe the possible realization of the magnetic DCDW in future HIC experiments at high baryon densities, due to its effect in the attenuation of certain light frequencies.

\section{Conclusions}
Cold and dense QCD in the magnetic DCDW phase exhibits a richness of macroscopically observable quantum topological effects that can be relevant for the HIC physics and neutron stars. It is rather fortunate that we will not have to wait too long to test the realization of these topological phases in the experiment. Future HIC experiments will certainly generate strong magnetic and electric fields in their off-central collisions and will open a much more sensitive window to look into a very challenging region of QCD. For example, the Compressed Baryonic Matter (CBM) at FAIR \cite{1607.01487}  have been designed to run at unprecedented interaction rates to provide high-precision measures of observables in the high baryon density region. That is why it is so timing and relevant to carry out detailed theoretical investigations of all potential observables of the magnetic DCDW phase. 

We hope that the findings presented here will serve to stimulate quantitative studies to identify signatures of the anomalous effects that will allow to probe the realization of this dense QCD phase both in neutron stars and in HIC experiments. 
\section{Acknowledgment}
The results presented in this paper were based on work done in collaboration with Efrain J Ferrer.

\end{document}